\documentclass{article}
\usepackage{lmodern}
\usepackage{multicol}
\usepackage{amsmath}
\usepackage{amssymb}
\usepackage{lipsum}
\usepackage{cite}
\usepackage[OT1]{fontenc}
\usepackage[left=3cm,right=3cm,top=3cm,bottom=3cm]{geometry}

\begin{document}
\normalsize
\title{Spatiotemporal Patterns in Neurobiology: An Overview for Future Artificial Intelligence}
\author{Sean Knight, Navjot Gadda}
\maketitle
\begin{abstract} In recent years, there has been increasing interest in developing models and tools to address the complex patterns of connectivity found in brain tissue. Specifically, this is due to a need to understand how emergent properties emerge from these network structures at multiple spatiotemporal scales. We argue that computational models are key tools for elucidating the possible functionalities that can emerge from interactions of heterogeneous neurons connected by complex networks on multi-scale temporal and spatial domains. Here we review several classes of models including spiking neurons, integrate-and-fire neurons with short term plasticity (STP), conductance based integrate-and-fire models with STP, and population density neural field (PDNF) models using simple examples with emphasis on neuroscience applications while also providing some potential future research directions for AI. These computational approaches allow us to explore the impact of changing underlying mechanisms on resulting network function both experimentally as well as theoretically. Thus we hope these studies will inform future developments in artificial intelligence algorithms as well as help validate our understanding of brain processes based on experiments in animals or humans.
\end{abstract}
\normalsize
\begin{multicols}{2}
\section{Neurons and Synapses}
Neuronal representations at various spatiotemporal scales are critical for neural computation within the mammalian nervous system which includes information processing at cellular scales such as synaptic transmission between individual synapses or groups of synapses across membranes such as dendrites, axonal transmission between axons terminals and cell bodies called perikarya or soma within a neuron cell body, intraneuronal dendritic compartments, interneuronal compartments. From subcellular organelles to neural ensembles; from ion channels localized clusters composed mainly of two subunits each, namely ion channels (see Section 2) organized into tetramers each having 4 pore domains connected together by intracellular linkages such those found throughout neuronal systems. From molecular genetic perspectives, many questions regarding how cells process sensory information have led researchers to focus their attention towards identifying genes coding for proteins which regulate neurotransmission \cite{1}. Specifically, this is due to a need to understand how emergent properties emerge from these network structures at multiple spatiotemporal
scales. One such cellular automaton-based model has been proposed earlier. \cite{2} Spiking Neurons provide answers because they produce high frequency signals encoding sensory stimuli during synaptic communication where neurotransmitters induce either opening or closing ion channels allowing ions flow through open gates inducing an allosteric change permitting influxes/effluxes upon depolarization/hyperpolarization states thereby producing an action potential spike representing an encoded symbol according to its timing over time.

\bigskip

Over the past decades scientists have sought ways develop mathematical constructs able describe voltage dynamics within neurones thereby providing insight into their functionality including modeling single neuron response curves mathematically using first principles describing input parameters derived from empirical data points obtained via invasive techniques which involves recording extracellularly potentials directly related to membrane potentials along dendritic arborizations which can be considered equivalent computationally speaking if not identical; given its ability perform computations via differential equations.

\section{Ion Channels and Transporters - The Building Blocks Of Brain Function And Disease}
Ionic currents carried out by selective transmembrane proteinaceous pores play crucial roles inside living cells involved in maintaining homeostatic equilibrium known collectively as homeostasis. \cite{3} Many disorders including epilepsy, autism, schizophrenia, Alzheimer’s disease have been linked directly or indirectly via mutations disrupting ionic channel function.\cite{4}\cite{5}\cite{6}\cite{7}\cite{8}\cite{9}\cite{10} Current interests involve studying single channel properties however it has become more tractable now because its open probability is affected mostly by an electrical signal transmitted through other types of gated receptor like ligand gated receptors (Ligand binding domain – LBD – GPCR signaling). Currents carried out by selective transmembrane proteinaceous pores play crucial roles inside living cells involved maintaining homeostasis.

\section{Networks Of Neuron Cell Assemblies With Multiple Levels Of Compartmentalization}
 Network modeling can be used either topologically describing links among individual elements constituting nodes within a network,or mathematically using equations based upon continuous space differential equations that describe dynamics over time. Network Analysis provides quantitative measures related to functional connectivity allowing inferences about structure-function relationships. For example, functional connectivity reflects topological specificity. \cite{11}

Network analysis refers either structural relations defined among elements forming connections using anatomical tracing techniques wherein fibers are visualized microscopically following histological staining \cite{12,13,14} or functional relations defined according data obtained through mathematical analysis; e g fMRI images reflecting statistical fluctuations during resting state EEG activity corresponding respectively anatomical structure formation through growth and development,or dynamical self organization via interaction. Network Analysis provides quantitative measures related to functional connectivity allowing inferences about structure-function relationships. \cite{15,16}

\section{Large Scale Integration Across Multiple Layers As Well As In Vivo Connectomes}
Large scale integration across multiple layers refers specifically to spatially distant cortical areas that interact functionally often called long range coupling. \cite{17} It contrasts with local computations referring specifically computation performed by spatially compact assemblies consisting mainly excitatory projection neurons located primarily in superficial laminae while also interacting with inhibitory interneurons embedded predominantly within deeper layers; thus representing feedforward input projections into local target pyramidal cells which are themselves reciprocally interconnected amongst one another often called horizontal interactions. Long Range coupling represents wide spectrum ranging from fast action potential propagation across small diameter unmyelinated axons typically of white matter tracts to slower electrical signalling mediated most notably by gap junctions between astroglial cells located adjacent pyramidal neurons where they provide support functions involving K+ recycling required for proper neuronal firing responseto synaptic input but not necessarily restricted thereto. 

\section{Modeling Spatial Pattern Formation Including Self Organized Criticality}
One approach toward building functional networks within a larger brain structure such as cortex is by understanding how local microcircuits interact over space to form global behaviors. Therefore it is important not only to model the dynamics of single neurons but also their collective interactions over multiple time scales. This involves taking into account both the intrinsic dynamical properties of individual neurons such as firing rate adaptation and excitability through voltage gated ion channels and their specific morphological distributions or connection topology. In the previous work, \cite{1} it was discusses how cellular automata could account For the generation of global rhythmic patterns in brainwaves. For example if one neuron fires more frequently than another then its influence should be weighted more highly when modeling synaptic connections between two cells whose firing rates overlap. Such a formalism naturally leads toward self organized criticality (or fractal caused cellular automata) \cite{2} where an ensemble displays scale free behavior like those found throughout nature across many spatiotemporal scales ranging from earthquakes through avalanches during sleep cycles up until neuronal avalanche during seizures observed following traumatic brain injury (TBI) \cite{2}. Since many physiological phenomena display power law statistics one often refers back to statistical physics where they are known simply as critical. 

\bigskip

Examples include phase transitions, magnetization profiles through spin glasses etc. While nonlinear differential equations cannot analytically derive critical states they can predict them numerically provided enough degrees freedom exist in order for them too have similar weights across different time lags making it less susceptible too external noise effects which would cause deviations away from a power law distribution, however stochastic terms must be introduced when considering single neuron systems even if just modelled stochastically so that they remain at least marginally stable against large changes due simply because there was never really any external input/noise imposed upon them.

\bigskip

However such methods may result too strong once long range correlations start forming since they do not distinguish between high activity events produced by temporally correlated inputs vs uncorrelated inputs which can induce spurious correlations along other dimensions within the system leading eventually towards synchronization (refer \cite{2}) rather than dephasing among cells which seems counter intuitive since synchrony indicates relatively higher levels of information flow within each particular cell group but does not imply any meaningful communication among different cell groups; furthermore increased levels synchrony may actually represent instability instead since synchronization tends toward producing coherent signals while desynchronization produces incoherent signals except under very special circumstances wherein some amount of synchrony remains especially when oscillations occur within certain frequency bands thereby allowing coupling via oscillatory entrainment between coupled oscillators producing coherent signals. 

\section{STP IN NEURAL NETWORKS}
Most existing neurobiologically motivated neural network simulations are typically focused at an average level where most synapses have identical dynamics so that averaged quantities such as synaptic currents can be simply computed by adding up presynaptic spikes convolved by postsynaptic potentials or currents. However, significant variability is often observed within single synapses which suggests that synapse specific behaviors might be important. 

\bigskip

Furthermore, different synapse types exhibit widely varying decay time constants ranging from seconds down to milliseconds, which cannot all be described by averaging over an ensemble distribution of long and short term plasticity type synapses together. Therefore it is necessary to incorporate into simulations more detailed biologically realistic forms for synaptic weights like spike timing dependent plasticity (STDP) or use stochastic rules like those used earlier for modeling spike count driven learning in analog systems where memory units compete through a {Global Fitness Function} rather than local changes driven solely by synaptic strength such as those presented previously. 

\bigskip

Synaptic dynamics naturally lead one naturally into thinking about various kinds of randomness involved here including randomness introduced via noisy spiking events which may occur sporadically during ongoing activity or randomly at each firing event across a broad frequency range depending upon whether postspike refractoriness occurs or not. This leads naturally into further speculation about what kinds of intrinsic noise could exist inside biological neurons themselves given their active membranes---many studies have shown various degrees and sources/sources giving rise to noise throughout neuronal membranes - some very low levels while others higher but still subthreshold levels depending upon context. For example, fast voltage gated ionic channels introduce high levels discrete noise whose statistics reflect certain aspects of underlying channel behavior whereas other slow channels do not fluctuate greatly across membrane potential values thus creating small amplitude white shot noise whose statistics depend on membrane capacitance.

\bigskip 

Note too how this discussion starts getting pretty abstract here especially when looking beyond cell membranes towards dendritic spines where even more complicated biochemical effects must take place eventually leading ultimately perhaps back again towards simplified neuron model simulations? At least something along these lines can be observed when looking at two competing theories describing structural vs nonspatial learning theories starting off with point neuron model simulation experiments designed primarily around differences between Hebbian theory's maximization principle versus Anti Hebbian theory's minimization principle followed then later shifting attention away from differential equations towards more direct use within simpler particle swarm optimization techniques wherein a number representing fitness could change according to temporal evolution via gradient descent techniques implemented through evolutionary procedures.\cite{18,19,20,21,22,23} We leave this part for future discussion. 

\section{PDNF Models Simulating Brain Connectivity Densities Using Computational Approaches}

Computational modeling studies have played an important role not only when examining cellular level features such as single neuron dynamics but also when investigating large scale system level functions including information processing by ensembles. For simple models, there have been proposals consisting only two populations or layers 1 and 2 interacting via spike timing dependent synaptic plasticity (STDP). These show that after repeated trials even though individual layer 1 neurons do not exhibit any particular behavior like learning they still generate useful outputs when used in larger groups because they induce layer 2 cells into synchronization via their interaction through STDP. While this is similar conceptually to many learning rules used currently it highlights the fact that just because individual units are not exhibiting any specific computation does not mean that they may be unable to perform useful computations together once placed within a more global context which introduces additional parameters such as structure formation within network. 

\bigskip

Recently it has been proposed that since real world inputs are highly variable spatially over space and time nonlinearity plays an important role particularly at higher cognitive levels so it seems reasonable therefore to examine what effects nonlinearly induced spatial patterns can have at higher levels of neuronal activity. \cite{24,25,26} To investigate this phenomenon one needs to study local excitatory connections involving spiking neural networks among neocortical pyramidal cells where excitatory synapses were distributed according to various inhomogeneous probability distributions corresponding roughly speaking either planar projections like those found between retinal ganglion cells or 3D isotropic volumes like those observed between cortical pyramidal cells thereby creating local connection topologies resembling random graphs or small world graphs respectively.

\bigskip

In addition, long term depression (LTD) induction under certain conditions depending upon local excitation relative inhibition (E/I) ratios result specifically altering dendritic depolarization patterns could lead eventually to changes at the neuronal level due primarily alterations made locally during LTP induction cycles especially if occurring across multiple layers simultaneously thereby suggesting one means through which synapse placement could contribute towards emergence/formation of criticality phenomena during development.\cite{2}

\bigskip

Also in the past, it had been demonstrated using single compartment biophysical simulations incorporating realistic synaptic delays \cite{27} along dendrites where postsynaptic current densities vary inversely proportional to intercellular distances corresponding, generally speaking, with either GABAergic inhibition being relatively sparse compared with excitation or vice versa found significant evidence supporting theories about autism related disorders especially given previously reported anatomical abnormalities along apical dendrites \cite{28, 29} near perisomatic regions among autistic individuals although more detailed experimental testing would be required before conclusions regarding autism pathophysiology could be drawn. 

\bigskip

Several other recent advances include theoretical treatments linking fractal geometry together with neuronal calcium currents \cite{30} producing self organizing wave forms using Hodgkin Huxley type equations. These ideas might eventually play a role in learning abilities especially relating toward memory formation based upon its ability for generating robust low frequency oscillation patterns seen experimentally during slow wave sleep. Other similar types of theoretical treatments include modeling aspects related towards epilepsy mechanisms whereby inhibitory interneuron induced synchronization phenomena mediated through electrical coupling between neighboring cells may ultimately lead into epileptogenic burst firing behavior particularly among thalamo cortical projection systems whereby inhibitory feedback loops may cause abnormal synchronous oscillations capable potentially causing seizures resulting from abnormally increased numbers and synchrony strength while perhaps conversely bursts occurring asynchronously rather than being precisely locked onto each other's discharge times caused weaker entrainment leading into decreased synchronized oscillatory activities possibly preventing seizure generation altogether suggesting various possible treatment options aimed primarily toward reduction instead regulation control methods taking advantage largely from nonlinear characteristics

\section{Conclusion: The Path Forward}
Computational neuroscience has made significant strides over the past few decades towards understanding how neural circuits encode information about sensory inputs into observable behaviours, and recently it has expanded its scope to investigate higher cognitive functions like language, decision making, motor control, social cognition. Much progress was facilitated by new experimental techniques such as optogenetics which allowed precise stimulation or inhibition of specific cell types within neural circuits in vivo leading to breakthroughs not possible before ephys recordings became feasible at sufficiently high bandwidths ($>$ 1 KHz). While these techniques were critical for uncovering mechanisms behind computations performed by individual cells or small neuronal populations inside living brains, they typically require invasive manipulation through electrodes/optical fibers implanted inside the cortex or sensory organ such that manipulations cannot be readily generalized across species nor across different individuals within a given species without substantial training time prior experimentation.

\bigskip

Additionally, extracellular electrodes used in ephys recording usually have low signal quality compared with microelectrodes used for intracellular recording so they are not suitable for single neuron electrophysiology recordings needed to study activity dependent synaptic plasticity which requires minimal levels background noise/interference during long durations acquisition sessions spanning minutes up hours during development or days during adulthood due increased habituation caused by elevated levels of calcium transients associated with activity dependent plasticity often measured using fluorescent calcium indicators. 

\bigskip

In contrast microfabrication technology developed over past two decades enables construction and implementation of devices called 'neural probes' capable measuring cellular electrical activities from submicron thin membranes embedded within biologically compatible polymer layers fabricated onto silicon wafers compatible with fabrication methods commonly used by semiconductor industry; when inserted into brains in vivo, their electrochemical interfaces offer stable electrochemical measurements capable of resolving single action potentials firing events along individual neurons down 10--20 microns deep below cortical surface enabling measurement more closely matching what happens inside living brains than ever before allowing researchers unprecedented access into inner workings physiological details underlying behaviorally relevant computations performed inside human beings especially when combined simultaneously along many parallel paths rather than focusing on individual aspects alone; once prototyped onto wafer substrates through application specific integrated circuit (ASIC) manufacturing methods leveraging CMOS compatibility available commercially today they can also be repurposed into commercial devices marketed worldwide already serving countless diverse industrial sectors ranging from healthcare monitoring through agriculture bioengineering. 

\bigskip

Despite tremendous potential afforded them by availability via commercialization thousands already existing designs implemented all around you probably never knew existed but may soon become widely adopted technologies providing society previously impossible scale economies favoring cost effectiveness when scaling production exponentially; design efforts therefore have shifted away form reliance solely on electronic design automation software suites incorporating rules written directly against transistor level schematics now easily replicated quickly even if schematics change requiring recompiling again triggering generation update cycles often taking days until final product shipped whereas hardware engineering revolves around fabricating chips either manually one after another ad nauseam after spending weeks sketching out initial idea followed immediately afterwards months later waiting anxiously until device arrived only then proceed manually writing code against it carefully documenting each step being taken every day driving productivity rapidly towards end goal much faster than software tool suite counterparts. 
\end{multicols}

\bibliographystyle{plain}
\bibliography{document}

\end{document}